\documentclass{llncs}

\usepackage[latin1]{inputenc}
\usepackage[english]{babel}
\usepackage{epstopdf}
\usepackage{graphicx,subfigure}
\usepackage{array}
\usepackage{amsmath,amsfonts,amssymb}

\usepackage[printonlyused]{acronym}
\acrodef{ASP}{Answer Set Programming}
\acrodef{PSN}{Protein Signaling Network}
\acrodef{PKN}{Prior Knowledge Network}
\acrodef{PSLM}{Protein Signaling Logic Model}
\acrodef{DNF}{Disjunctive Normal Form}
\acrodef{ILP}{Integer Linear Programming}
\acrodef{GA}{Genetic Algorithm}
\acrodef{CSP}{Constraint Satisfaction Problem}
\acrodef{MSE}{Mean Square Error}
\acrodef{LSSA}{Logical Steady State Analysis}
\acrodef{LSS}{Logical Steady State}

\title{Revisiting the Training of Logic Models of Protein Signaling Networks with a Formal Approach based on Answer Set Programming}

\author{
	Santiago~Videla\inst{1}\inst{2}\inst{7}
 	\and
	Carito~Guziolowski\inst{3}\inst{7}
	\and
	Federica~Eduati\inst{4}\inst{5}
	\and
	Sven~Thiele\inst{2}\inst{1}\inst{6}
	\and
	Niels~Grabe\inst{3}
	\and
	Julio~Saez-Rodriguez\inst{5}§
	\and
	Anne~Siegel\inst{1}\inst{2}§}
\institute{
CNRS, UMR 6074 IRISA, Campus de Beaulieu, 35042 Rennes cedex, France.
\and
INRIA, Centre Rennes-Bretagne-Atlantique, Projet Dyliss, Campus de Beaulieu, 35042 Rennes cedex, France.
\and
University Heidelberg, National Center for Tumor Diseases, Hamamatsu TIGA Center, Im Neuenheimer Feld 267, 69120 Heidelberg, Germany.
\and
Department of Information Engineering, University of Padova, Padova, 31050, Italy.
\and
European Bioinformatics Institute (EMBL-EBI) Wellcome Trust Genome Campus, Cambridge CB10 1SD, UK.
\and
University of Potsdam, Institute for Computer Science, Germany.
\and
These authors contributed equally to this work
}
\begin{document}

\maketitle

\renewcommand{\thefootnote}{§}
	\footnotetext{Corresponding authors: saezrodriguez@ebi.ac.uk, anne.siegel@irisa.fr}
\renewcommand{\thefootnote}{\arabic{footnote}}\setcounter{footnote}{0}

\textbf{To appear at Lecture Notes in Computer Science, Springer}
\begin{abstract}
A fundamental question in systems biology is the construction and training to data of mathematical models. Logic formalisms have become very popular to model signaling networks because their simplicity allows us to model large systems encompassing hundreds of proteins. An approach to train (Boolean) logic models to high-throughput phospho-proteomics data was recently introduced and solved using optimization heuristics based on stochastic methods. Here we demonstrate how this problem can be solved using \ac{ASP}, a declarative problem solving paradigm, in which a problem is encoded as a logical program such that its answer sets represent solutions to the problem. \ac{ASP} has significant improvements over heuristic methods in terms of efficiency and scalability, it guarantees global optimality of solutions as well as provides a complete set of solutions. We illustrate the application of \ac{ASP} with \textit{in silico} cases based on realistic networks and data.
\end{abstract}


\section{Introduction}
Cells perceive extracellular information via receptors that trigger
signaling pathways that transmit this information and process
it. Among other effects, these pathways regulate gene expression
(transcriptional regulation),
thereby defining the response of the cell to the information sensed in
its environment. Over decades of biological research we have gathered
large amount of information about these pathways. Nowadays, 
there exist public repositories such as Pathways Commons \cite{Cerami2011} and Pathways Interaction Database \cite{Schaefer2009} that
contain curated regulatory knowledge, from which signed and oriented graphs can be automatically retrieved \cite{Zinovyev2007,Guziolowski2012}. 
These signed-oriented graphs represent molecular interactions inside the cell 
at the levels of signal transduction and (to a lower extent) of transcriptional regulation.
Their edges describe causal events, which in the case of signal
transduction are  related to the molecular events triggered by cellular receptors.
These networks are derived from vast generic knowledge concerning different cell types and
they represent a useful starting point to generate predictive models for cellular events.

Phospho-proteomics assays \cite{Palmisiano2010} are a recent form of
high-throughput or 'omic' data.
They measure the level of phosphorylation (correlated with protein-activity) 
of up to hundreds of proteins at the same moment in a particular biological system~\cite{Terfve2012}.
Most cellular key processes, including proliferation, migration, and
cell cycle, are  ultimatelly controlled by these protein-activity
modifications. Thus, measurement of phosphorylation of key proteins
under appropriate conditions  (experimental designs), such as  stimulating or perturbing the system in different ways, can provide
useful insights of cellular control.

Computational methods to infer and analyze signaling networks from high-throughput phospho-proteomics data are less mature than for
transcriptional data, which has been available for much longer time~\cite{Terfve2012}.
In particular, the infererence of gene regulatory networks from transcriptomics data is now an established field (see
\cite{Bansal2007,Hecker2009} for a review). In comparison to
transcriptomics, data is harder to obtain in (phospho) proteomics, but
prior knowledge about the networks is much more abundant, and available
in public resources as mentioned above.

An approach to integrate the prior knowledge existing in databases with the specific insight provided by phospho-proteomics data was
recently introduced and implemented in the tool CellNOpt
(CellNetOptimizer; www.cellnopt.org) \cite{Saez-Rodriguez2009}.
CellNOpt uses stochastic optimization algorithms (in particular, a genetic algorithm), to find the Boolean logic model compatible
that can best describe the existing data. While CellNOpt has proved able to train networks of realistic size, it suffers from the
lack of guarantee of optimum intrinsic of stochastic search
methods. Furthermore, it scales poorly since the search space (and
thus the computational time) increases exponentially with the network size.

In this paper, we propose a novel method to solve the optimization
problem posed in  \cite{Saez-Rodriguez2009} that overcomes its limitations. Our approach trains generic networks 
based on experimental measures equally as CellNOpt in order to obtain a complete set of global optimal networks specific to the experimental data.
This family of optimal networks could be regarded as an explanatory model that is specific to a particular cell type and condition; 
from these models it should be possible to derive new, more accurate biological insights.
To illustrate our approach we used a generic \ac{PKN} related to
signaling events upon stimulation of cellular receptors in hepatocytes,
and trained this network with \textit{in silico} simulated phospho-proteomics
data. This network was used as a benchmark for  network inference in
the  context of the DREAM (Dialogues for Reverse Engineering
Assessment of Methods; www.the-dream-project.org) Predictive Signaling Network Challenge
\cite{Prill2011}.

The proposed solution encodes the optimization problem in Answer Set Programming (ASP)~\cite{baral03}.
ASP is a declarative problem solving paradigm from the field of logic programming.
Distributed under the GNU General Public Licence,
it offers highly efficient inference engines based on Boolean constraint solving technology~\cite{gekanesc07a}.
ASP allows for solving search problems from the complexity class NP and
 with the use of disjunctive logic programs from the class $\Sigma^{P}_{2}$. 
Moreover, modern ASP tools allow handling complex preferences and multi-criteria optimization,
 guaranteeing the global optimum by reasoning over the complete solution space.
  
Our results show significant improvements, concerning computation time and completeness in the search
of optimal models, in comparison with CellNOpt. We note that similar
features can be obtained by formulation of the problem as an integer
linear optimization problem~\cite{Mitsos2009}.
The perspectives of this work go towards the exploration of the complete 
space of optimal solutions in order to 
identify properties such as the robustness of optimal models, and
relate them to the quality of the obtained
predictions. 

\section{Formalization}\label{sec:formalization}
The biological problem that we tackle in this work is essentially a combinatorial optimization problem over the possible logic models representing a given \ac{PKN}. In this section, first we introduce the graphical representation of logic models by giving a simple example that motivates our formalization. Then, we give a formal definition for the inputs of the problem, we formally define a \ac{PSLM} and we show how predictions are made for a given model. Finally, we define an objective function used for the optimization over the space of possible logic models.

\subsection{Motivation}
\begin{figure}[ht!]
\centering
\subfigure[]{
  \label{fig:example-PKN}
  \includegraphics[scale=0.25]{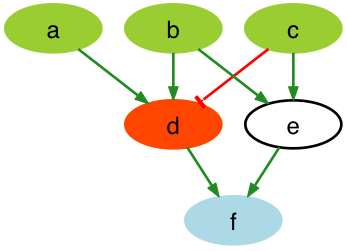}
}
\subfigure[]{
  \label{fig:example-model}
  \includegraphics[scale=0.25]{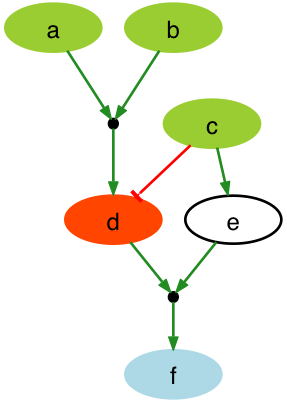}
}

\caption{\scriptsize {\bf Hypergraph representation of Logic Models}. The green and red edges correspond to activations and inhibitions, respectively. Green nodes represent ligands that can be experimentally stimulated. Red nodes represent those species that can be inhibited by using a drug. Blue nodes represent those species that can be measured by using an antibody. White nodes are neither measured, nor manipulated.
{\bf (a)} 
A toy \ac{PKN} as a directed and signed graph.
{\bf (b)} An arbitrary Logic Model compatible with the \ac{PKN} shown in (a). Black filled circles represent AND gates, whereas
multiple edges arriving to one node model OR gates.
}\label{fig:example-hyper}
\end{figure}

The functional relationships of biological networks, such as PKNs, cannot be captured using only a graph~\cite{Klamt2006,Saez-Rodriguez2009}. If, for example, two proteins (nodes) A and B have a positive effect on a third one C (encoded in a graph as A $\rightarrow$ C, and B $\rightarrow$ C), is not clear if either A or B can active C, or if both are required (logic OR and AND gate, respectively). To  represent such complex (logical) relations between nodes and offer a formal representation of cellular networks, hypergraphs can be used. Since hypergraphs were already described and used to represent logic models of protein signaling networks  in~\cite{Klamt2006,Saez-Rodriguez2007,Saez-Rodriguez2009,Christensen:2009gr,Klamt2009}, here we adopt the same formalism and we simply give an example to introduce this representation. For more details, we refer the reader to the cited literature. 
\begin{example}
	\small
Given the toy \ac{PKN} described in Fig.~\ref{fig:example-PKN}, an arbitrary compatible logic model is given by the following set of formulas $\{d = (a \land b) \lor \neg{c};e = c;f = d \land e\}$. Moreover, a representation of this logic model is given in Fig.~\ref{fig:example-model} as a signed and directed hypergraph. Note that each conjunction clause gives place to a different hyperedge having as its source all the present literals in the clause.
\end{example}

Informally, the training of logic models to phospho-proteomics data consist in finding the hypergraph(s) compatible with a given \ac{PKN} that best explains the data (using some criteria). Even though a graphical representation is quite intuitive and has been widely used in the literature, it is not the most appropriate way to give a formal and clear formulation of this problem. Thus, in what follows we give a formalization based on propositional logic and in the rest of this work, whenever convenient, we will refer interchangeably to Protein Signaling Logic Models as hypergraphs and to conjunctive clauses as hyperedges.

\subsection{Problem Inputs}
We identify three inputs to the problem: a \acf{PKN}, a set of experimental conditions or perturbations and for each of them, a set of experimental observations or measurements. For the sake of simplicity, in this work we have considered only \acp{PKN} with no feedback loops. They account for the main mechanisms of transmission of information in signaling pathways, but do not include feedback mechanisms that are typically responsible for the switching off of signals once the transmission has taken place~\cite{Saez-Rodriguez2009,Terfve2012}. In what follows, we give a mathematical definition for each of these inputs.
\begin{definition}[Prior Knowledge Network]\label{def:PKN}
A \ac{PKN} is a signed, acyclic, and directed graph $(V,E,\sigma)$ with $E \subseteq V\times V$ the set of directed edges, $\sigma \subseteq E \times \lbrace 1,-1 \rbrace$ the signs of the edges , and $V=S \cup K \cup R \cup U$, the set of vertices where $S$ are the \textit{stimulus} (inputs), $K$ are the \textit{inhibitors} (knock-outs), $R$ are the \textit{readouts} (outputs) and $U$ are neither measured, nor manipulated. Moreover, the subsets $S,K,R,U$ are all mutually disjoint except for $K$ and $R$.
\end{definition}

Note that in the previous definition, $\sigma$ is defined as a relation and not as a function since it could be the case where both signs are present between two vertices. This is even more likely to happen when a \ac{PKN}, either extracted from the literature or from one of the mentioned databases, is compressed as described in~\cite{Saez-Rodriguez2009} in order to remove most of the nodes that are neither measured, nor manipulated during the experiments. Also note that the subset of nodes $K$ correspond to those proteins (e.g. kinases) that can be forced to be inactive (inhibited) by various experimental tools such as small-molecule drugs, antibodies or RNAi.
\begin{definition}[Experimental condition]\label{def:experiment}
Given a \ac{PKN} $(V,E,\sigma)$ an experimental condition over $(V,E,\sigma)$ is a function $\varepsilon: S\cup K \subseteq V \rightarrow \lbrace 0,1 \rbrace$ such that if $v \in S$, then $\varepsilon(v)=1$ (resp. $0$) means that the stimuli $v$ is present (resp. absent), while if $v \in K$, then $\varepsilon(v)=1$ (resp. $0$) means that the inhibitor for $v$ is absent and therefore $v$ is not inhibited (resp. the inhibitor for $v$ is present and therefore $v$ is inhibited). 
\end{definition}
\begin{definition}[Experimental observation]\label{def:observation}
Given a \ac{PKN} $(V,E,\sigma)$ and an experimental condition $\varepsilon$ over $(V,E,\sigma)$, an experimental observation under $\varepsilon$ is a function $\theta: R(\varepsilon) \subseteq R \rightarrow \lbrace 0,1 \rbrace$ such that $R(\varepsilon)$ denotes the set of observed readouts under $\varepsilon$ and if $v \in R(\varepsilon)$, $\theta(v) = 1$ (resp. $0$) means that the readout $v$ is present (resp. absent) under $\varepsilon$.
\end{definition}

Since the phospho-proteomics data used here represents an average across a population of cells, each of which may contain a different number of proteins in active or inactive  (1 or 0) state, the values are continuous.  Thus, we have to discretize the experimental data somehow in order to fit the previous definition. A simple but yet effective approach is to use a threshold $t=0.5$ such that values greater than $t$ are set to 1, while values lower that $t$ are set to 0. Other approaches could also be used but, since in this paper we work with discrete \textit{in silico} data, we left this discussion for a future work. Indeed, this paper focuses on comparing the performance of training and formal approaches to the optimization problem, for which \textit{in silico} datasets appear more relevant.

\subsection{Protein Signaling Logic Models}
Here we state the combinatorial problem as a \acf{CSP} in order to have a clear and formal definition of a \acf{PSLM} as a solution to this problem. Recall that a \ac{CSP} is defined by a set of variables $X$, a domain of values $D$, and a set of constraints or properties to be satisfied. A solution to the problem is a function $e: X\rightarrow D$ that satisfies all constraints~\cite{Tsang1993}.

Next, we define two properties that we use later as the constraints of the \ac{CSP} formulation. The first property defines for a given \ac{PKN}, the conditions that must be satisfied by a logical formula in order to define the truth value of any node. For example, if we look the Fig.~\ref{fig:example-hyper} is quite clear that the hypergraph in (b) is not just some arbitrary hypergraph, but instead is strongly related to the graph in (a). This relation is captured by the following definition.
\begin{definition}[\ac{PKN} evidence property]\label{def:PKN-evidence}
Given a PKN $(V,E,\sigma)$ and $v \in V$, a logical formula $\varphi$ in \ac{DNF} has an evidence in $(V,E,\sigma)$ with respect to $v$ if and only if for every propositional variable $w$ that occurs positively (resp. negatively) in $\varphi$, it exists an edge $(w,v) \in E$ and $((w, v), 1) \in \sigma$ (resp. $((w, v),-1) \in \sigma$).
\end{definition}

The second property identifies those logical formulas in \ac{DNF} for which exist some equivalent but simpler formula. For example, for two literals $X$ and $Y$, it is easy to see that $X \lor (X \land Y) \equiv X$. In such case we say that $X \lor (X \land Y)$ is redundant since, as we will see later, we are interested in minimizing the complexity of the logic models. This concept was previously introduced in~\cite{Saez-Rodriguez2009} as a way to reduce the search space of all possible logical formulas.
\begin{definition}[Redundancy property]\label{def:redundancy}
Given a logical formula $\varphi$ in \ac{DNF}, with $\varphi = \bigvee_{j\ge1}{ c_{j}}$ where each $c_{j}$ is a conjunction clause, $\varphi$ is a redundant formula if and only if for some $k,l \ge 1$ with $k \ne l$ and some logical conjunction $r$ it holds that $c_{k} = c_{l} \land r$.
\end{definition}

Now, based on the general form of a \ac{CSP} given above, and the properties defined in (\ref{def:PKN-evidence}) and (\ref{def:redundancy}), we define a \ac{PSLM} as follows.
\begin{definition}[Protein Signaling Logic Model]\label{def:pslm}
Given a \ac{PKN} $(V,E,\sigma)$ with $V \setminus S = \lbrace v_{1},\dots,v_{m}\rbrace$ for some $m \ge 1$, let the set of variables $X = \lbrace \psi_{v_{1}},\dots, \psi_{v_{m}} \rbrace$ and the domain of values $D$ given by all the formulas in \ac{DNF} having $V$ as the set of propositional variables. Then, a function $e: X \rightarrow D$  defines a compatible Protein Signaling Logic Model $\mathcal{B}$ if it holds for $i=1,\dots,m$ that $e(\psi_{v_{i}})$ has an evidence in $(V,E,\sigma)$ with respect to $v_{i}$. Moreover, if it also holds for $i=1,\dots,m$ that $e(\psi_{v_{i}})$ is not redundant, then we say that $\mathcal{B}$ is a non redundant logic model.
\end{definition}
\begin{example}
		\small
	Given the \ac{PKN} in Fig.~\ref{fig:example-PKN} the function that defines the logic model in Fig.~\ref{fig:example-model} is given by: 
	\[
	e(\psi_{v}) = \left\{
	\begin{array}{l l}
	    (a \land b) \lor \neg{c} & \quad \text{if } v = d\\
	    c & \quad \text{if } v = e\\
		(d \land e) & \quad \text{if } v = f
	  \end{array} \right.
	\]
	for $\psi_{v} \in \lbrace \psi_{d}, \psi_{e}, \psi_{f}\rbrace$. Note that in every case, each formula satisfies both properties: PKN evidence (Definition~\ref{def:PKN-evidence}) and Non-redundancy (Definition~\ref{def:redundancy}).
\end{example}

\subsection{Predictive Logic Model}
A given \acf{PSLM} describes only the static structure of a Boolean network. 
Even though Boolean networks are either synchronous or asynchronous, in any case
 the set of \acp{LSS} is the same. 
Therefore, and since we focus on a \ac{LSSA} which offers a number of applications for studying functional aspects
 in cellular interactions networks~\cite{Klamt2006},
 choosing between synchronous and asynchronous is not relevant in this work. 
Moreover, we do not need to compute all possible \acp{LSS},
 but only the one that can be reached from a given initial state. 
Note that the existence of a unique \ac{LSS} is guaranteed by the assumption of no feedbacks loops in the given \ac{PKN}. 
Next, we describe how we compute this \ac{LSS} in terms of satisfiability of a particular logical formula. 
This is based on the formalization given by~\cite{Haus2009} for a related problem named IFFSAT.

Let $(V,E,\sigma)$ a \ac{PKN} and $\mathcal{B}$ a compatible \ac{PSLM}. 
Recall that $\mathcal{B}$ is defined by a function from every non-stimuli in $(V,E,\sigma)$ to a \ac{DNF} formula satisfying the \ac{PKN} evidence property defined in Definition \ref{def:PKN-evidence}. 

First, we define a logical formula $\mathcal{R}$ representing the regulation of every non-stimuli or non-inhibitor node in $(V,E,\sigma)$.
\begin{equation}
\mathcal{R} = \bigwedge_{v \in V \setminus (S\cup K)} (\mathcal{B}(v) \iff v)
\end{equation}

Note that we use $(\mathcal{B}(v) \iff v)$ instead of just $(\mathcal{B}(v) \Rightarrow v)$ 
 to enforce that every activation must have a ``cause'' within the model. 
Next, we define two logical formulas $\mathcal{S}$ and $\mathcal{K}$ in order to fix the values of \textit{stimulus} and
 the values or regulations of \textit{inhibitors} in $(V,E,\sigma)$ under a given experimental condition $\varepsilon$. 
\begin{equation}
\mathcal{S} = \bigwedge_{v \in S}
\begin{cases}
	v & \quad \text{if } \varepsilon(v)=1 \\
	\neg v & \quad \text{if } \varepsilon(v) = 0
\end{cases}
\quad\quad\quad
\mathcal{K} = \bigwedge_{v \in K}
\begin{cases}
	\mathcal{B}(v) \iff v & \quad \text{if } \varepsilon(v) = 1 \\
	\neg v & \quad \text{if } \varepsilon(v) = 0
\end{cases}
\end{equation}

Thereafter, we look for the truth assignment such that $\mathcal{R} \wedge \mathcal{S} \wedge \mathcal{K}$ 
evaluates to \textit{true} and which represents the \ac{LSS} of the network for the given initial conditions. 
The biological meaning behind this concept is that the input (\textit{stimuli}) signals are propagated through the network
 by using the faster reactions and after some time, the state of each protein will not change in the future. 
Thus, we say that the network is stabilized or that it has reached an steady state~\cite{Klamt2006}. 
Finally, we can define the model prediction under a given experimental condition as follows.
\begin{definition}[Model prediction]\label{def:prediction}
Given a \ac{PKN} $(V,E,\sigma)$, a \ac{PSLM} $\mathcal{B}$ compatible with $(V,E,\sigma)$ and a experimental condition $\varepsilon$ over $(V,E,\sigma)$, the prediction made by $\mathcal{B}$ under the experimental condition $\varepsilon$ is given by the truth assignment $\rho:V \rightarrow \lbrace 0,1 \rbrace$ such that $\mathcal{R} \wedge \mathcal{S} \wedge \mathcal{K}$ evaluates to true.
\end{definition}

Note that without the assumption of no feedbacks loops in the given \ac{PKN}, the existence of multiple steady states or cycle attractors should be considered. Then, in order to guarantee that $\rho$ is well defined, new constraints should be added to the \ac{CSP} instance defined in~(\ref{def:pslm}), but this is left as a future work.
\begin{example}
		\small
	Let $\varepsilon: \{a,b,c,d\} \rightarrow \{0,1\}$ an experimental condition over the \ac{PKN} given in Fig.~\ref{fig:example-PKN} defined by:
	\[
	\varepsilon(v) = \left\{
	\begin{array}{l l}
	    1 & \quad \text{if } v \in \{a,b,c\}\\
	    0 & \quad \text{if } v = d
	  \end{array} \right.
	\]
\end{example}
That is, $a,b,c$ are stimulated while $d$ is inhibited. Then, the prediction made by the \ac{PSLM} given in Fig.~\ref{fig:example-model} under $\varepsilon$, is given by the truth assignment such that the formula
\[
((d \land e) \iff f) \land (c \iff e) \land a \land b \land c \land \neg{d}
\]
evaluates to $true$. Thus, $e$ is assigned to 1 and $f$ is assigned to 0.

\subsection{Objective function}
Given all the \acp{PSLM} compatible with a given \ac{PKN},
 our goal is to define an objective function in order to capture under different experimental conditions,
 the matching between the corresponding experimental observations and model predictions. 
To this end, we adopt and reformulate the objective function proposed in~\cite{Saez-Rodriguez2009} in terms of our formalization.
The objective function represents a balance between fitness of model to experimental data and model size using a free parameter chosen to maximize the predictive power of the model. 
Of course, other objective functions can be defined in the future, but here we focus on a comparison against one of the state of the art approaches and thus, we choose to use the same objective function.

Before going further, we define the size of a model as follows.
\begin{definition}[Size of Protein Signaling Logic Models]\label{def:model-size}
Given a \ac{PKN} $(V,E,\sigma)$ and a \ac{PSLM} $\mathcal{B}$ compatible with $(V,E,\sigma)$, the size of $\mathcal{B}$ is given by $|\mathcal{B}| = \sum_{v \in V \setminus S} |\mathcal{B}(v)|$ where $|\mathcal{B}(v)|$ denotes the canonical length of logical formulas.
\end{definition}
\begin{example}
		\small
	If we consider the \ac{PSLM} given in Fig.~\ref{fig:example-model}, its size is given by: $|(a \land b) \lor \neg{c}| + |c| + |(d \land e)| = 3 + 1 + 2 = 6$. This can be seen also as the size of the hypergraph, where each hyperedge is weighted by the number of source nodes and the size of the hypergraph is the sum of all weights.
\end{example}

Finally, we define the combinatorial optimization problem of learning \acp{PSLM} from experimental observations under several experimental conditions as follows.
\begin{definition}[Learning Protein Signaling Logic Models]\label{def:learning-pslm}
Given a \ac{PKN} $(V,E,\sigma)$, $n$ experimental conditions $\varepsilon_{1},\dots,\varepsilon_{n}$ 
 and $n$ experimental observations $\theta_{1},\dots,\theta_{n}$ with each $\theta_{i}$ defined under $\varepsilon_{i}$,
 for a given \ac{PSLM} $\mathcal{B}$ compatible with $(V,E,\sigma)$,
 and $n$ model predictions $\rho_{1},\dots,\rho_{n}$ over $\mathcal{B}$ with each $\rho_{i}$ defined under $\varepsilon_{i}$, we want to minimize
\begin{equation}
\label{equ:objective}
	\Theta(\mathcal{B}) = \Theta_{f}(\mathcal{B}) + \alpha \times \Theta_{s}(\mathcal{B})
\end{equation}
where $\Theta_{f}(\mathcal{B}) = \frac{1}{n_{o}} \times \sum_{i=1}^{n}\sum_{v \in R(\varepsilon_{i})} (\theta_{i}(v) - \rho_{i}(v))^{2}$ such that $n_{o}$, the total number of output measures, is given by $\sum_{i=1}^{n} |R(\varepsilon_{i})|$ and $\Theta_{s}(\mathcal{B}) = \frac{1}{b} \times |\mathcal{B}|$ such that $b$ is the size of the union of all \acp{PSLM} compatible with $(V,E,\sigma)$.
\end{definition}

\section{Methods}\label{sec:methods}
In this section we describe the methods used to perform a comparison between our ASP-based approach and the one presented in~\cite{Saez-Rodriguez2009}. First we provide the \ac{ASP} implementation that we used to run the experiments and then, we describe the method proposed to systematically generate \textit{in silico} study cases based on realistic networks and data.

\subsection{\ac{ASP} implementation}\label{sec:encoding}
Our goal is to provide an \ac{ASP} solution for learning \acp{PSLM} from experimental observations under several experimental conditions (Definition~\ref{def:learning-pslm}). Here we provide a logic program representation of the problem described in Section~\ref{sec:formalization} in the input language of the \ac{ASP} grounder \textit{gringo} \cite{Gebser:2009bp}. After describing the format of any input instance, we show how we generate non-redundant candidate solutions having an evidence in the given \ac{PKN}, then we describe how model predictions are made and finally, we show the minimization of the objective function.

\subsubsection{Input instance}
We start by describing the input instance for the \ac{PKN} given by $(V,E,\sigma)$, the experimental conditions $\mathcal{E} = \varepsilon_{1},\dots,\varepsilon_{n}$ and the experimental observations $\mathcal{O} = \theta_{1},\dots,\theta_{n}$.
\small
\begin{equation}
	\begin{split}
		\mathcal{G}((V,E,\sigma),\mathcal{E},\mathcal{O}) =& 
		\lbrace vertex(v)\ |\ v \in V \rbrace \\
		\cup & \lbrace edge(u,v,s)\ |\ (u,v) \in E, ((u,v),s) \in \sigma \rbrace \\
		\cup & \lbrace exp(i,v,s)\ |\ \varepsilon_{i}(v) = s, v \in S \cup K \rbrace \\
		\cup & \lbrace obs(i,v,s)\ |\ \theta_{i}(v) = s, v \in R(\varepsilon_{i}) \rbrace \\
		\cup & \lbrace nexp(n) \rbrace \\
		\cup & \lbrace stimuli(v)\ |\ v \in S \rbrace \\
		\cup & \lbrace inhibitor(v)\ |\ v \in K \rbrace \\
		\cup & \lbrace readout(v)\ |\ v \in R \rbrace
	\end{split}
\end{equation}
\normalsize
\subsubsection{Candidate solutions}
We follow a common methodology in \ac{ASP} known as ``guess and check'' where using non-deterministic constructs, we ``guess'' candidate solutions and then, using integrity constraints we ``check'' and eliminate invalid candidates. Since we are interested only on those logical formulas having an evidence in $(V,E,\sigma)$, first we generate all the possible conjunction clauses having such evidence by computing for every $v \in V$ all the possible subsets between the predecessors of $v$. This is done by the following rules.
\small
\begin{equation}\label{rule:subsets}
	\begin{split}
		subset(U,S,null,1,V) \leftarrow&\ edge(U,V,S). \\
		subset(U,S_{U},subset(W,S_{W},T),N+1,V) \leftarrow&\ subset(U,S_{U},null,1,V), \\
		&\ subset(W,S_{W},T,N,V), \\
		&\ vertex(U) < vertex(W).
	\end{split}
\end{equation}
\normalsize
The idea is to start with the singleton subsets containing only a single predecessor, and to create a bigger subset by recursively extending a singleton subset with any other subset until a fix point is reached. The first rule defines all the singleton subsets related to $V$. We represent the subsets here as linked lists where $U$, the first argument in the predicate $subset/5$, represents the head of the list (5 is the arity of the predicate). The second argument represents the sign of the edge from $U$ to $V$. The third argument represents the tail of the linked list ($null$ in case of a singleton). The fourth argument represents the subset cardinality, and the last argument keeps track of the target vertex. The head is here used as a identifier such that we can order all subsets. The second rule recursively extends a singular subset identified by head argument $U$ with any subset identified by $W$ as long as $U < W$. We exploit the order between the predicates $vertex/1$ to avoid different permutations of the same subsets.

The following rules define the inclusion relationship between these subsets of predecessors.
\small
\begin{equation}\label{rule:inclusion}
	\begin{split}
		in(U,S,subset(U,S,T)) \leftarrow&\ subset(U,S,T,N,V). \\
		in(W,S_{W},subset(U,S_{U},T)) \leftarrow&\ in(W,S_{W},T), \\&\ subset(U,S_{U},T,N,V).
	\end{split}
\end{equation}
\normalsize
The first rule declares that every subset contains its ``head'' element. The second rule declares that if $W$ is included in $T$, and if there is another subset having $T$ as its ``tail'', then $W$ is also included in it.

Since each subset generated by the rules in~(\ref{rule:subsets}) represents a possible conjunction clause, we can generate all possible logical formulas in \ac{DNF} by considering each subset as either present or absent.
\small
\begin{equation}\label{rule:clause}
	\begin{split}
		\lbrace clause(subset(U,S,T),N,V) \rbrace \leftarrow&\ subset(U,S,T,N,V). \\
		\leftarrow&\ clause(C_{1},N,V), clause(C_{2},M,V), \\
		&\ C_{1} \ne C_{2}, in(U,S,C_{2}) : in(U,S,C_{1}).
	\end{split}
\end{equation}
\normalsize
The first rule is a choice rule that declares the non-deterministic generation of predicates $clause/3$ from a subset. A clause represents the conjunction of all the elements included in the subset. The second rule declares an integrity constraint to avoid the generation of redundant logical formulas by using the predicates generated in~(\ref{rule:inclusion}).

\subsubsection{Model predictions}

Next, we show the representation for the input signals propagation. For each experiment, first the truth values for stimuli and inhibited nodes are fixed and then, truth values are propagated to all nodes by exploiting the fact that in order to assign \textit{true} to any node, it is enough that one conjunction clause over it evaluates to \textit{true}.
\small
\begin{equation}
\label{rule:fixed}
	\begin{split}
		fixed(E,V) \leftarrow&\ nexp(N), E=1..N, stimuli(V).\\
		fixed(E,V) \leftarrow&\ inhibitor(V), exp(E,V,0).\\
		active(E,V) \leftarrow&\ exp(E,V,1), stimuli(V).\\
		inactive(E,V) \leftarrow&\ exp(E,V,0).
	\end{split}
\end{equation}
\normalsize
The first and second rules simply declare which nodes have fixed truth values because they are either an input node,
 or an inhibited node in a particular experiment. 
Thereafter, the third and fourth rules declare the truth assignments that are given by the experimental condition.

The following rules model the signal propagation in every experiment.
\small
\begin{equation}
\label{rule:propagation}
	\begin{split}
		active(E,V) \leftarrow&\ nexp(N), E=1\dots N, \\
		&\ clause(C,M,V), \textit{not } fixed(E,V),  \\
		&\ active(E,U) : in(U,1,C), \\&\ inactive(E,U) : in(U,-1,C). \\
		inactive(E,V) \leftarrow&\ vertex(V), nexp(N), E=1\dots N, \\
		&\ \textit{not } fixed(E,V), \textit{not } active(E,V).
	\end{split}
\end{equation}
\normalsize
The first rule declares that for each experiment,
 if there is at least one conjunction clause having all its positive literals assigned to \textit{true} and
 all its negated literals assigned to \textit{false}, 
 then the complete clause evaluates to \textit{true}.
While the second rule declares that every node that is not assigned to \textit{true},
 it is assigned by default to \textit{false}.

\subsubsection{Optimization}
Finally, we show the declaration of the objective function. In Section~\ref{sec:formalization} we defined the objective function $\Theta$, but since \ac{ASP} can only minimize integer functions, we transformed $\Theta$ into $\Theta_{\text{int}}$ trying to lose as less information as possible. To this end, if we assume that the free parameter $\alpha = \frac{N}{D}$ for some $N,D \in \mathbb{N}$, multiplying $\Theta$ by $N \times \frac{1}{\alpha} \times n_{o} \times b$ we define $\Theta_{\text{int}}$ as follows.
\small
\begin{equation}\label{equ:objective-int}
	\begin{split}
	\Theta_{\text{int}}(\mathcal{B}) =&\ D \times n_{o} \times b \times \Theta(\mathcal{B}) \\
	=&\ D \times b \times \sum_{i=1}^{n}\sum_{v \in R(\varepsilon_{i})} (\theta_{i}(v) - \rho_{i}(v))^{2} + N \times n_{o} \times |\mathcal{B}|
	\end{split}
\end{equation}
\normalsize
This new (integer) objective function is represented as follows in our ASP encoding.
\small
\begin{equation}\label{rule:minimization-aux}
\begin{array}{lcl}
  \#const\ npenalty=1.  & & \\
  \#const\ dpenalty=1000.   & &\\
  penalty\_N(npenalty). & &\\
  penalty\_D(dpenalty). & &\\
  b(B) & \leftarrow&\ B = [subset(\_,\_,N,\_) = N].\\
  n_{o}(E)& \leftarrow&\ E = [obs(\_,\_,\_)]. \\
  mismatch(E,V)& \leftarrow&\ obs(E,V,0), active(E,V).\\
  mismatch(E,V) &\leftarrow&\ obs(E,V,1), inactive(E,V).\\
\end{array}
\end{equation}
\normalsize
The rules in~(\ref{rule:minimization-aux}) declare the predicates that we need to give a representation of $\Theta_{\text{int}}$. First, we declare two predicates to represent the free parameter $\alpha$ as a fraction of integers. 
Then, we use a weighted sum to declare the size of the union of all logic models and we count the number of single experimental observations. The two last rules declare in which cases a model prediction does not match the corresponding output measurement.

Last but not least, we require the minimization of the (integer) objective function $\Theta_{\text{int}}$ simply by using the $\#minimize$ directive.
\small
\begin{equation}
	\begin{split}
		\#minimize[&\ mismatch(\_,\_) : b(B) : penalty\_D(PD) = B \times PD, \\
		&\ clause(\_,N,\_) : n_{o}(E) : penalty\_N(PN) = E\times PN \times N].
	\end{split}
\end{equation}
\normalsize

\subsection{Benchmark datasets}\label{sec:benchmarks}
We wanted to compare the \ac{ASP} approach with CellNOpt and analyze the scalability of the methods.
Also we wanted to determine how the inference of the network is influenced by specific parameters of the problem.
For this purpose, we generated meaningful benchmarks that covered a broad range of these influential parameters.

\subsubsection{Middle and large-scale benchmark datasets}
\begin{figure}[ht!]
\centering

\subfigure[]{
  \label{fig:pkn}
  \includegraphics[width=4cm]{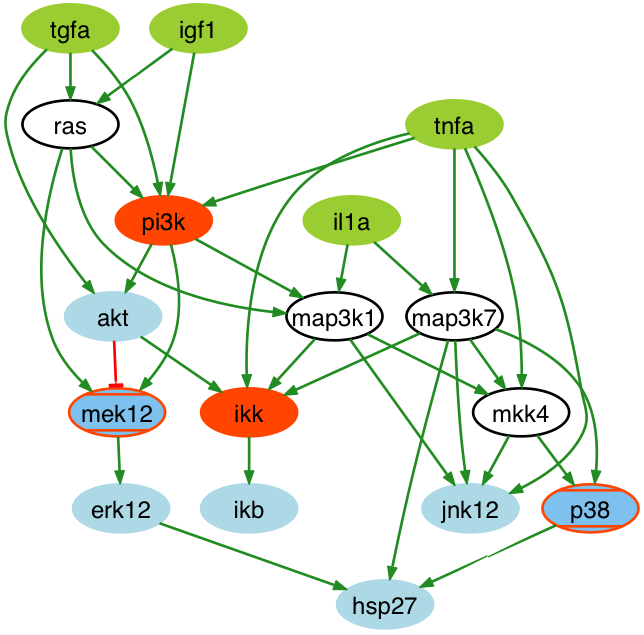}
}
\subfigure[]{
  \label{fig:benchmark}
  \includegraphics[width=4cm]{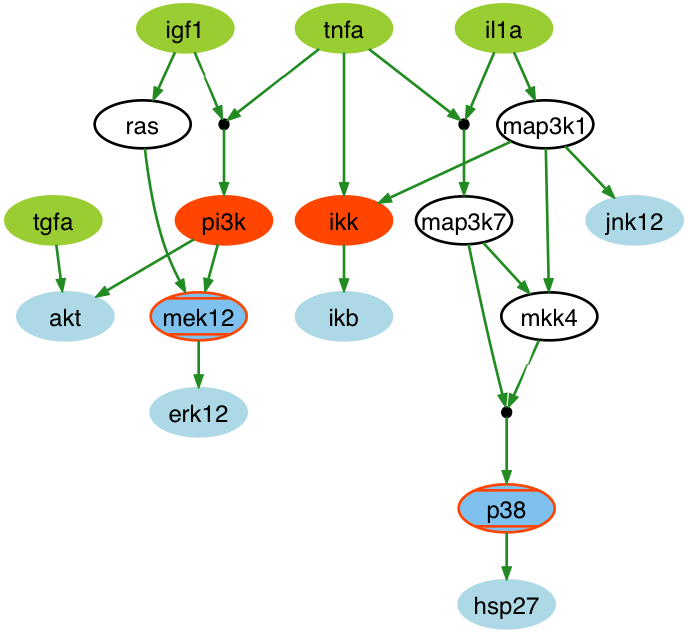}
}
\subfigure[]{
\label{fig:optimal}
\includegraphics[width=5cm]{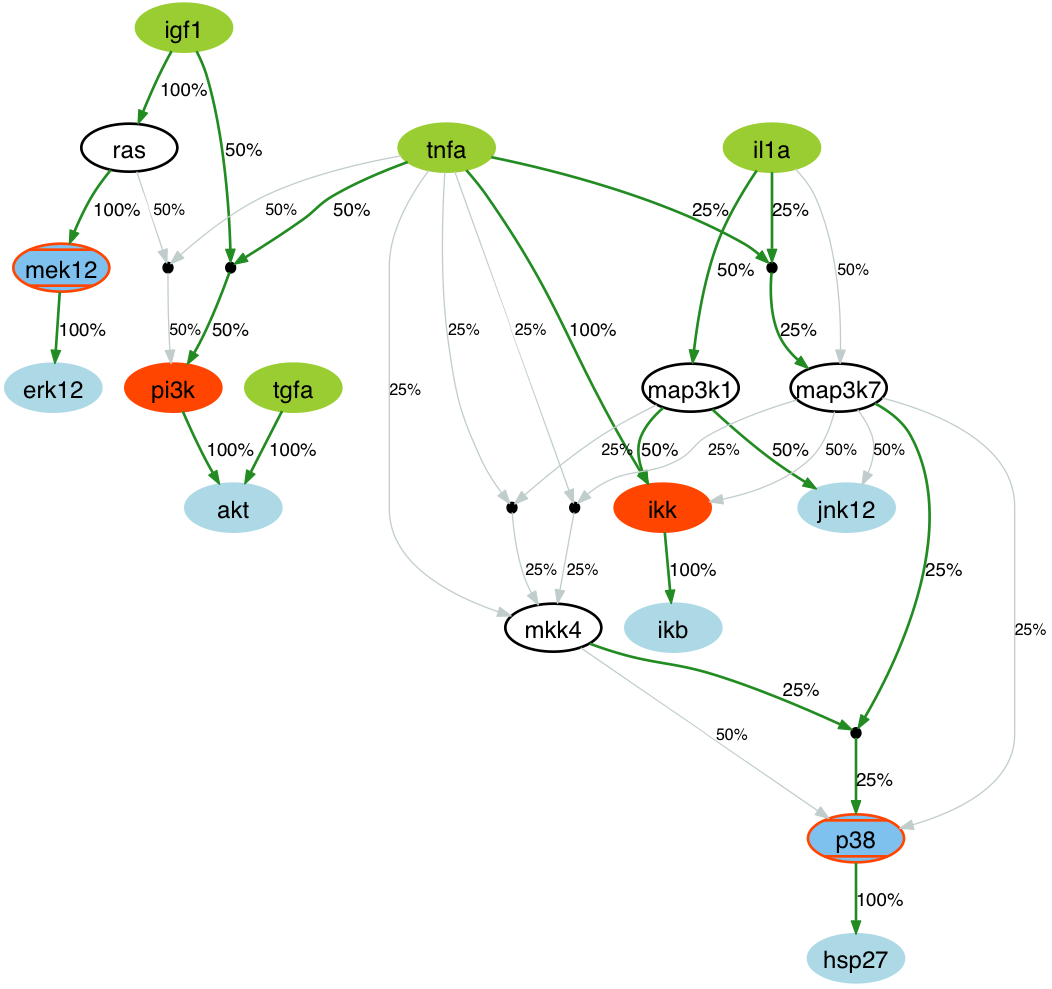}
}
\caption{\scriptsize {\bf Input and outputs of a middle-scale optimization problem}.
{\bf (a)} A literature-derived \acf{PKN} of growth and inflammatory signaling. 
{\bf (b)} An hypergraph which is compatible with the PKN shown in (a). From this model we derive 240 output measures under 34 experimental conditions.
{\bf (c)} The \ac{ASP} optimization enumerated all the minimal \acp{PSLM} that predict the \textit{in silico} measures produced in (b) with no mismatches. The union of the 8 optimal models is shown with a specific edge encoding: edges are labeled according to their percentage of occurrence in all the 8 models. The thick green edges correspond to those edges that also appear in the hypergraph used to generate the \textit{in silico} datasets.
}\label{fig:toy}
\end{figure}

\begin{figure}[ht!]
\centering
\subfigure[]{
  \label{fig:pkn-liver}
  \includegraphics[width=4cm, height=5cm]{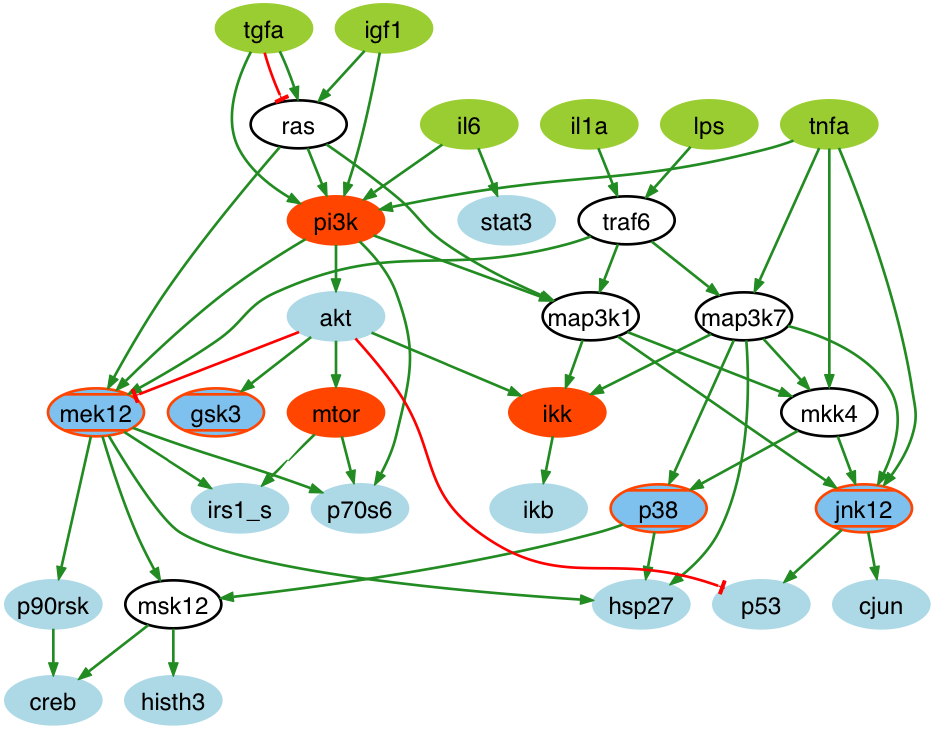}
}
\subfigure[]{
  \label{fig:benchmark-liver}
  \includegraphics[width=4cm,height=5cm]{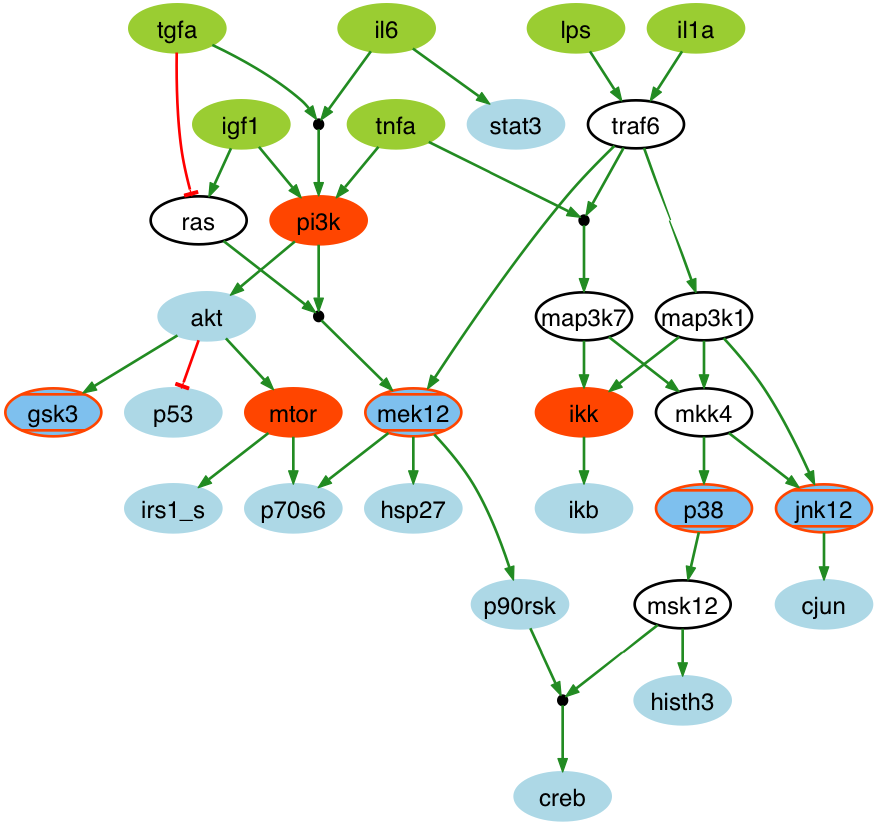}
}
\subfigure[]{
  \label{fig:optimal_large}
 \includegraphics[width=5cm]{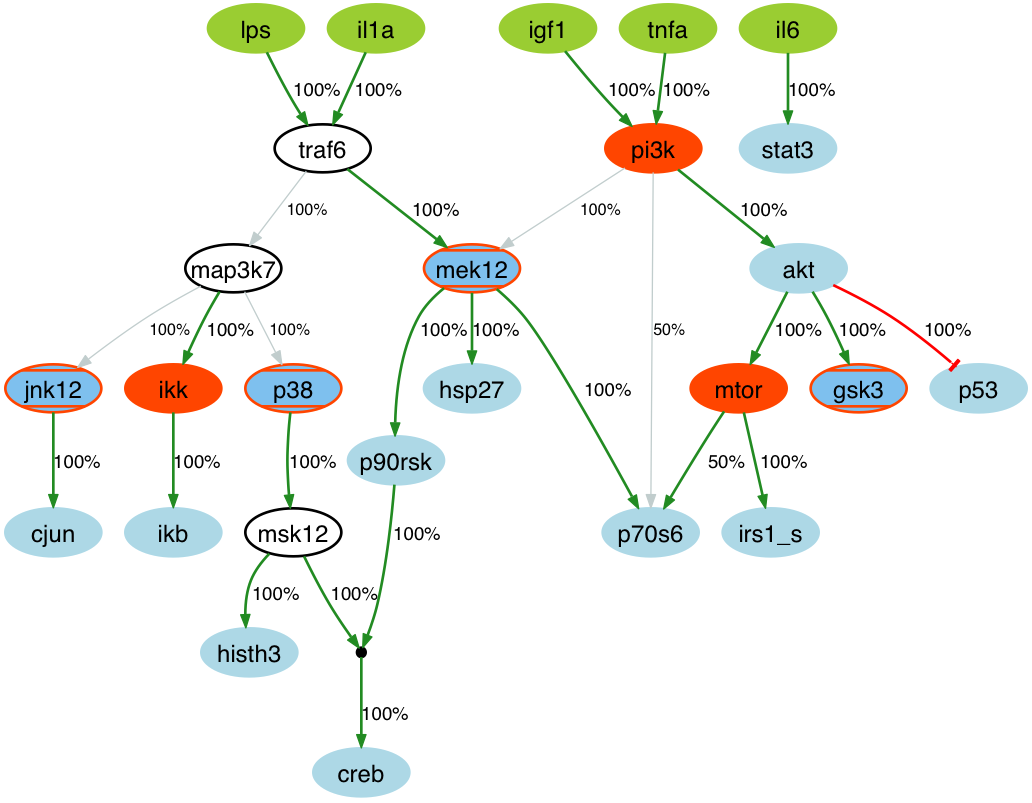}
}
\caption{\scriptsize {\bf Large-scale optimization problem}.
{\bf (a)}  A Large literature-derived \ac{PKN} of growth and inflammatory signaling obtained from \cite{Morris2011}.
{\bf (b)} An hypergraph which is compatible with the \ac{PKN} shown in (a). From this model we derive 840 output measures under 56 experimental conditions.
{\bf (c)}  Union of the two minimal \acp{PSLM} predicting the whole dataset with no mismatches.
}\label{fig:realNet}
\end{figure}

We constructed a middle (see Fig \ref{fig:toy}) and a large scale (see Fig \ref{fig:realNet}) optimization problem. Both \acp{PKN} were derived from literature and in each case we randomly selected compatible \acp{PSLM} or hypergraphs (middle: Fig.~\ref{fig:benchmark}, large: Fig.~\ref{fig:benchmark-liver}), from which we generated \textit{in silico} datasets under several experimental conditions giving place to different numbers of output measures. The main parameters used to compute the objective function for the optimization are shown in the Table~\ref{tab:middle-large}.

\begin{table}
  \small
  \caption{\scriptsize Middle and Large optimization problems}
  \centering
  \begin{tabular}{| c | c | c | >{\centering\arraybackslash}m{1.7cm} | >{\centering\arraybackslash}m{1.7cm} | >{\centering\arraybackslash}m{1.7cm} | >{\centering\arraybackslash}m{2cm} | >{\centering\arraybackslash}m{1.6cm} |}
	\hline
	Scale & Nodes & Edges & Compatible hyperedges & Size of union hypergraph & Selected hypergraph size & Experimental conditions & Output measures \\
	\hline
	Middle & 17 & 34 & 87 & 162 & 20 & 34 & 210 \\
	\hline
	Large & 30 & 53 & 130 & 247 & 37 & 56 & 840 \\
	\hline
  \end{tabular}
  \label{tab:middle-large}
\end{table}

\vspace{-1cm}

\subsubsection{Large set of benchmark datasets}
We relied on a literature derived \ac{PKN} for growth and inflammatory signaling~\cite{Prill2011} to derive compatible \acp{PSLM} and generate 240 benchmark datasets with \textit{in silico} observation data. Given the literature derived network $(V,E,\sigma)$ with $V=S \cup K \cup R \cup U$, we created 4 derivative networks $(V_i,E,\sigma), i = 1 \dots 4$ with $V_i=V$, $V_i= S \cup K_i \cup R_i \cup U_i$, $K_i \subseteq K$, $R_i \subseteq R$, and $U \subseteq U_i$. Each network differing in sets of inhibitors and readouts. For these networks we compressed ('bypassed) nodes that are neither measured, nor manipulated during the experiments, which were not affected by any perturbation, lay on terminal branches or in linear cascades, as described in~\cite{Saez-Rodriguez2009}, yielding to 4 compressed networks.

To investigate the influence of the size of the networks in the optimization both in terms of computational times and recovered edges, we randomly selected \acp{PSLM} of 3 different sizes (20, 25, or 30) for each compressed network. 
The size of each model was obtained as defined in Definition~\ref{def:model-size}. Then, 5 different \acp{PSLM} were generated for each compressed network and for each size, giving a total of 60 different models (20 of each size). We use these 60 models to run simulated experiments and generate \textit{in silico} experimental observations.

\begin{figure}
\centering
\includegraphics[scale=0.55]{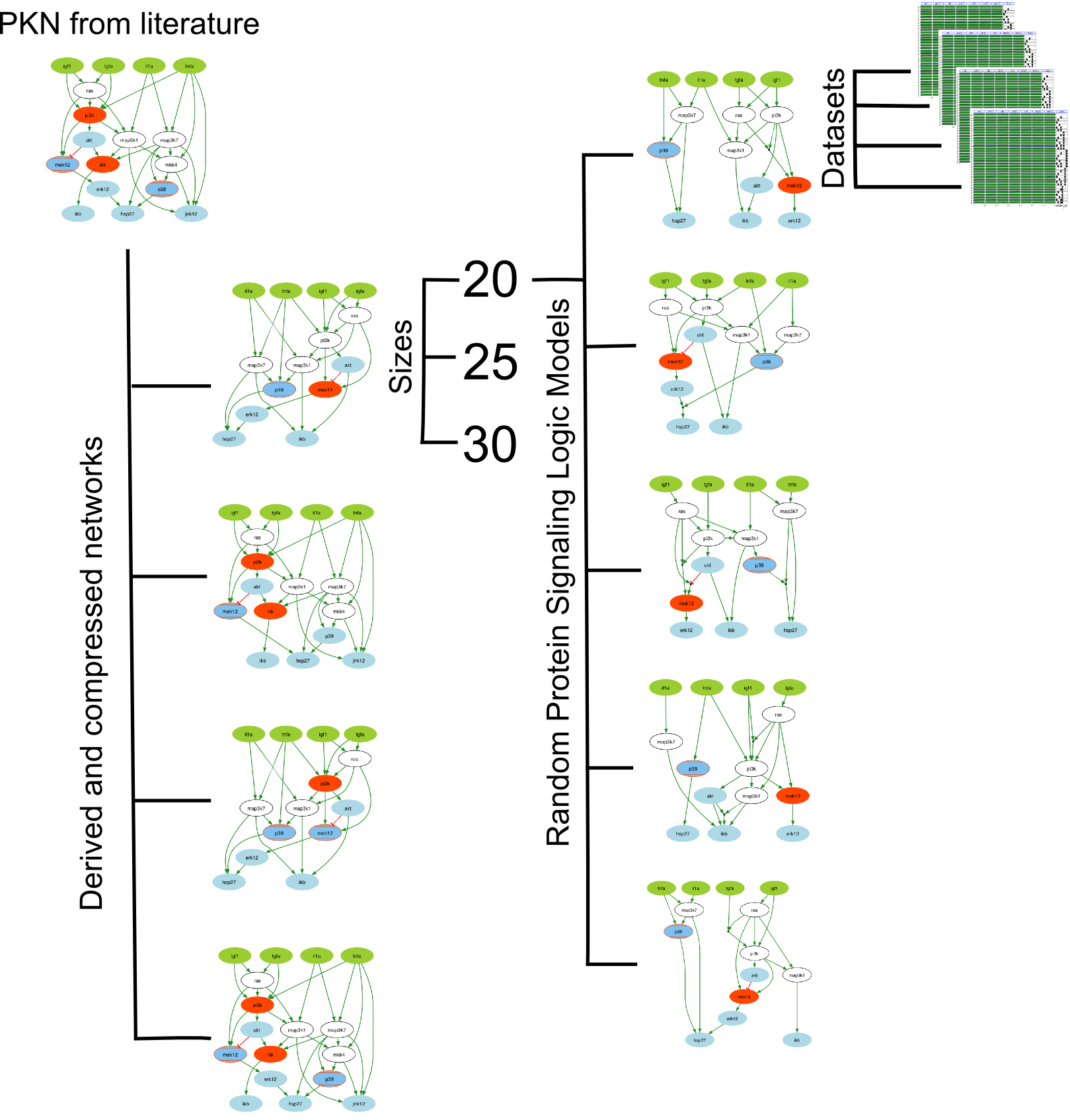}
\caption{\scriptsize Pipeline of the generation of the 240 benchmarks}
\label{fig:benchmarks-gen}
\end{figure}

Moreover, we wanted to investigate how the amount of experimental observation data influences the network inference.
Therefore, we generated 4 datasets $D_1\dots D_4$ of experimental observations for each model.
The first dataset $D_1$ contained only experimental observations from single-stimulus/single-inhibitor experiments. The other datasets $D_2 \dots D_4$ contained observations from multiple-stimuli/multiple-inhibitors experiments with 30, 50 and 60 experimental conditions respectively. The larger datasets always include the smaller datasets, such that $D_1 \subset D_2 \subset D_3 \subset D_4$. In total we generated 240 different datasets of 4 different sizes, generated from 60 different models of 3 different sizes. The whole method is illustrated in the Fig.~\ref{fig:benchmarks-gen}.

\section{Results and discussions}
First, we focused in finding minimal \acp{PSLM} compatible with the given \ac{PKN} and predicting the generated dataset for the middle (see Fig.~\ref{fig:toy}) and large-scale (see Fig.~\ref{fig:realNet}) benchmarks.
Second, general comparisons between our logical approach implemented in \ac{ASP} and the genetic algorithm  implemented in CellNOpt, were performed over the 240 datasets generated as described in Section \ref{sec:benchmarks}. 

\subsection{Enumeration of solutions to the optimization problems}
We used the \ac{ASP} implementation detailed in Section~\ref{sec:encoding} to identify the \acp{PSLM} compatible with the middle-scale \ac{PKN} (respectively the large-scale \ac{PKN}) having an optimal score with respect to the generated dataset.

Notice that in both cases, by the construction of the datasets, we knew that there exists a compatible \ac{PSLM} which predicts the whole datasets without mismatches. As a consequence, if $\alpha \in (0,1)$ (see Eq.~\ref{equ:objective}, Eq.~\ref{equ:objective-int}), the optimization problem is equivalent to find the logic models with perfect fit and minimal size.

In a first run, the \ac{ASP} implementation allowed us to compute the minimal score of the optimization problem.  Afterwards, we run the \ac{ASP} solver again to enumerate all the models having a score lower or equal than the minimal score. All together, we obtained a complete enumeration of all minimal models. Below, we show the results obtained using the \ac{ASP}-based approach to solve the middle and large optimization problems.
\begin{itemize}
\item{\textbf{Middle-scale}} The minimal score was computed in 0.06 seconds\footnote{All ASP computations were run in a MacBook Pro, Intel Core i7, 2.7 GHz and 4 GB of RAM using Gringo 3.0.3 and Clasp 2.0.5 versions.}. The enumeration took 0.03 seconds and found 8 global optimal Boolean models with size equal to 16. The union of the 8 optimal models found is shown in Fig. \ref{fig:optimal}.
\item{\textbf{Large-scale}} The minimal score was computed in 0.4 seconds. The enumeration took 0.07 seconds and found 2 global optimal Boolean models with size equal to 26. In Fig.~\ref{fig:optimal_large} we show the union of the 2 optimal models found.
\end{itemize}

Both optimization problems were also run with CellNOpt, based on its genetic algorithm (see Materials and Methods section in~\cite{Saez-Rodriguez2009}) performing generations over a population of 500 models\footnote{All CellNOpt computations were run in a cluster of 542 nodes, each with 32 GB of memory, and a total of 9000 cores using CellNOptR 1.0.0.}. 
\begin{itemize}
\item{\textbf{Middle-scale}} The optimization was run for 9.2 hours and the best score was reached after 7.2 hours (299 generations). During the optimization, 66 Boolean models with perfect fit were found, with sizes going from 16 to 24. Out of the 66 models, only 2 models were minimal (i.e. with size equal to 16).
\item{\textbf{Large-scale}} The optimization problem was run for 27.8 hours and the best score was reached after 24.5 hours (319 generations). During the optimization, 206 models with perfect fit were found, with sizes going from 27 to 36. Note that in this case, CellNOpt did not find any of the minimal models (i.e. with size equal to 26).
\end{itemize}

Our main conclusion here is as follows: in both cases, due to the use of \textit{in silico} data, models with perfect fit were exhibited by both approaches. The main advantage of the formal approach is to be able to explicitly compute the minimal score, allowing us to enumerate all models with this score in a very short time. Meanwhile, genetic algorithms are not able to exhibit this information and therefore cannot develop strategies to compute all minimal models. At the same time, this leads to the question about the biological relevance of optimal models and if it is possible to discriminate between them. A precise study of the biological pathways selected in each optimal model did not allow us to specifically favor one model according to biological evidences. That is why we choose to show the union of them in each case (Fig.~\ref{fig:optimal} and Fig.~\ref{fig:optimal_large}). 

Nonetheless, the \ac{ASP} search was strongly supported by the fact that there exists at least one model with perfect fit. This considerably reduces the optimization problem to the search of compatible models with minimal size by canceling the $\Theta_f$ term in Eq.~(\ref{equ:objective}). Performing optimizations over real data will induce that there will no more exist models with perfect fit, which may have a strong effect over the performance of our formal approach, while for genetic algorithms performances may probably be less affected by real data, but this will have to be studied. An interesting perspective is therefore to test the efficiency of these approaches in a real case experiment. 
\begin{figure}[ht!]
\centering
  \includegraphics[width=12cm]{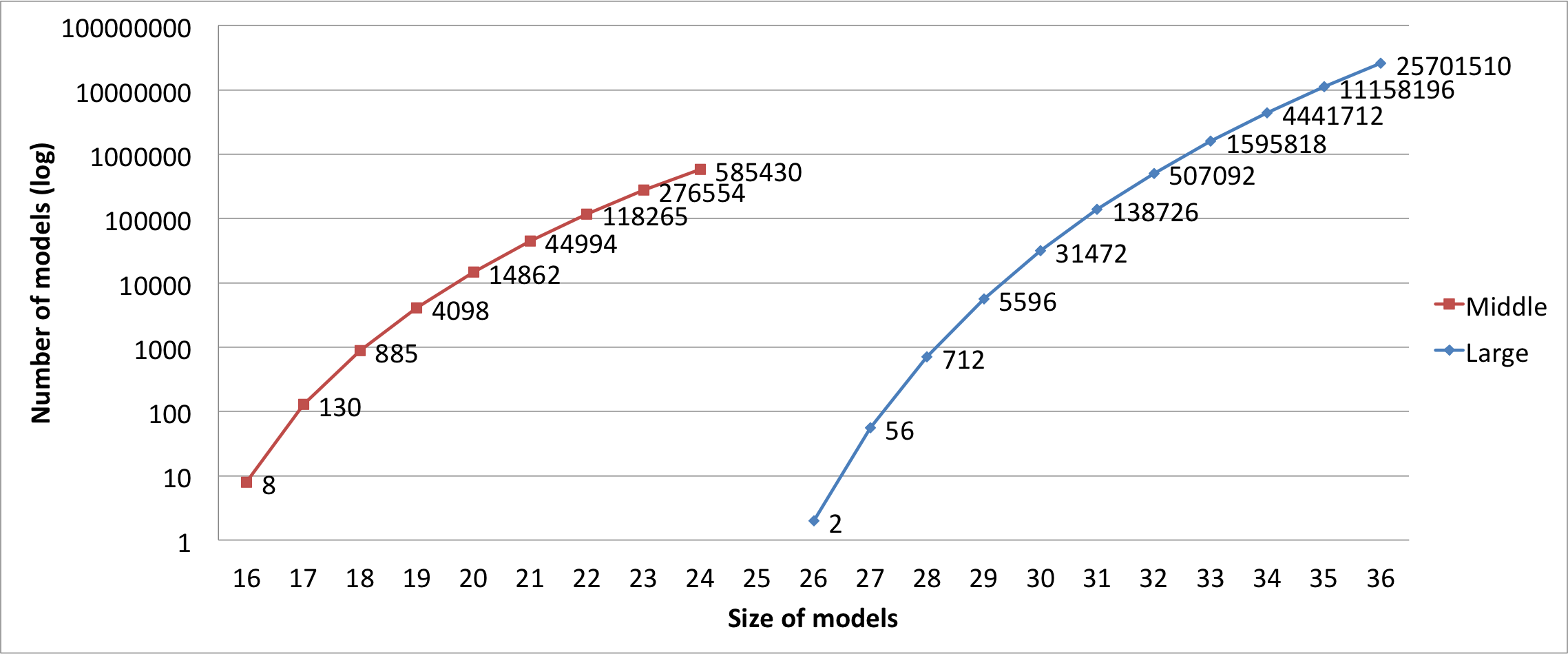}
\caption{\scriptsize {\bf   Number of suboptimal solutions  to the middle-scale (red curve) and large-scale (blue curve) optimization problems}.  Each curve describes the number of models with perfect fit with a given size, where the size ranges from its minimal value to the maximal size of models found by CellNOpt.}\label{fig:suboptimal}
\end{figure}
\subsection{Dependency to the model size}
To have a first view of the space of solutions, we investigated the role of the model size over the optimization process. Indeed, the optimization criteria 
moderates the choice of a model of minimal size -according to a parsimonious principle- by a free parameter related to the fitting between observations and predictions. (see Eq.~\ref{equ:objective}). However, as we mentioned above, in all our experiments we known that there exists at least one model which predicts the whole datasets without mismatches and thus, the optimization problem is focused on finding minimal models. Therefore strongly favoring the size of the model. To evaluate the impact of this for the middle and large optimization problems depicted in Fig.~ \ref{fig:toy} and Fig.~\ref{fig:realNet}, we used \ac{ASP} to enumerate all the models with perfect fit having their size less or equal to the size of the models found by CellNOpt. Results are depicted in Fig.  \ref{fig:suboptimal}, providing a first insight on the structure of the space of compatible \acp{PSLM} with perfect fit. It appears that the number of compatible \acp{PSLM} increases exponentially with the size of the model. Therefore, optimizing over the size criteria appears quite crucial. A prospective issue is to elucidate whether the topology of the space of suboptimal models informs about the biological relevance of minimal models.
\begin{figure}[hb!]
\centering
\includegraphics[width=12cm]{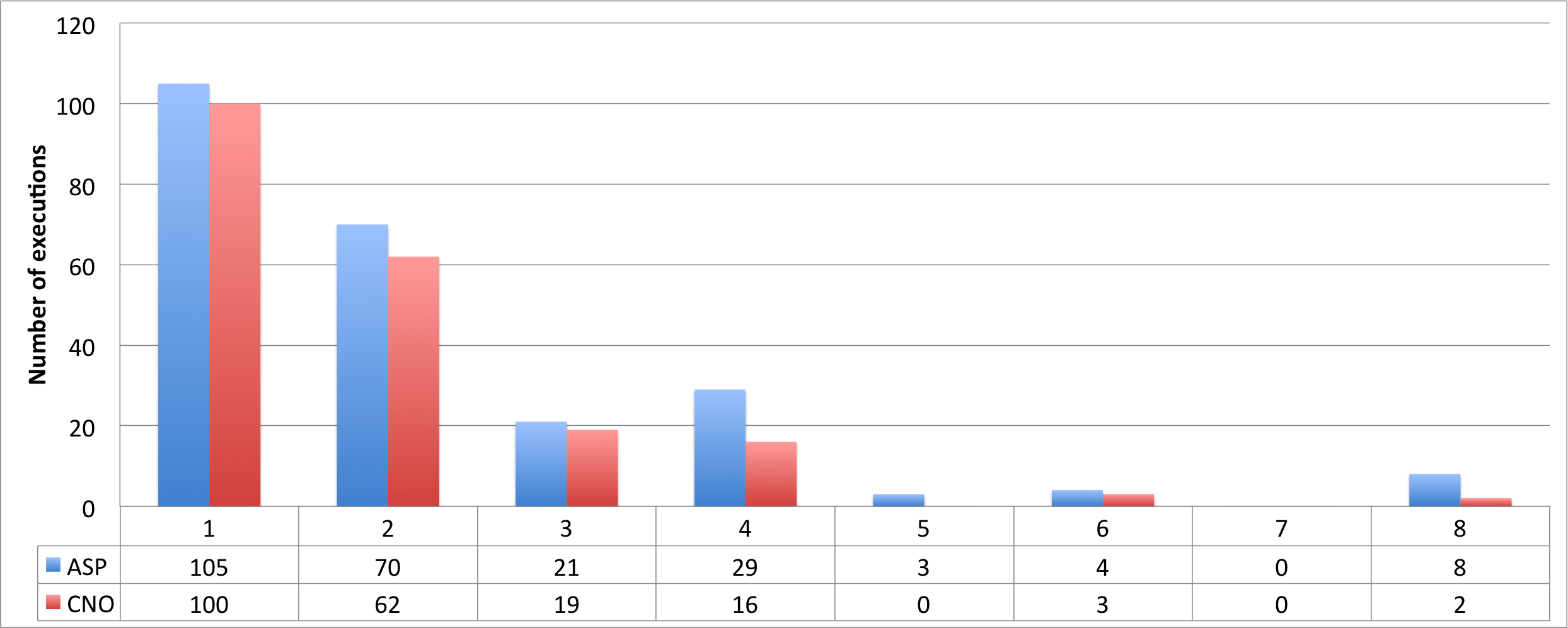}
\caption{\scriptsize {\bf Executions of \ac{ASP} and CellNOpt optimizations that found all global optimal models.} 
The total number of runs was of 240 in account of the in-silico data generated. The x-axis represents the
number of global optimal models that each problem had. The y-axis, shows the number of 
executions where \ac{ASP} and CellNOpt found the total number of global optimums.}
\label{fig:optimizations240}
\end{figure}

\subsection{Accuracy of predictions}
The study of the middle and large optimization problems evidenced that genetic algorithms may not find all minimal models. In order to elucidate whether this phenomenon is frequent, we used the 240 benchmark datasets generated with the method described in Section~\ref{sec:benchmarks}. In Fig.~\ref{fig:optimizations240} we show the number of executions of the optimization process for \ac{ASP} and CellNOpt where both approaches found the complete set of global optimal (minimal) models. Recall that \ac{ASP} ensures finding the complete set of global optimal models (blue bars in Fig.~\ref{fig:optimizations240}) while this is not the case for CellNOpt. We observed that in 202 executions out of 240 (84\%), \ac{ASP} and CellNOpt both found all the minimal models. This is particularly clear in the 105 executions with a single minimal model, which was found by CellNOpt in 95\% of executions. Nonetheless, in the 44 cases with more than 4 optimal models, CellNOpt found all optimal models in only 47\% cases.  More generally, as the number of minimal solutions to the optimization problem increases, the percentage of minimal solutions identified by CellNOpt decreases.
\begin{figure}[hb!]
\centering
\includegraphics[width=12cm]{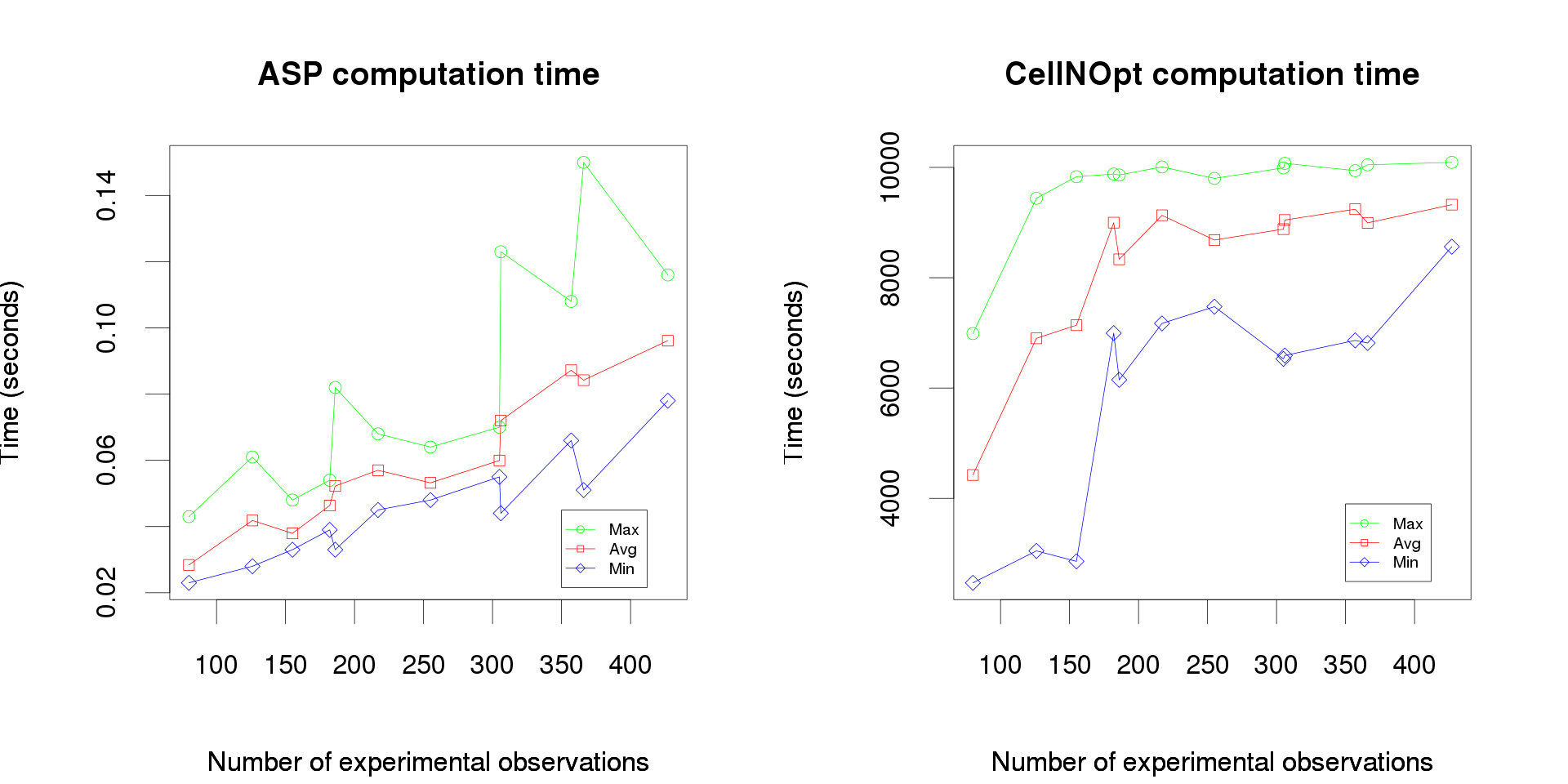}
\caption{\scriptsize{\bf Computation time of \ac{ASP} and CellNOpt with respect to the number of experimental observations.} 
The x-axis is the number of experimental observations: sum of number of readouts for each experimental condition.
The maximum, average, and minimum computation times are plotted in green, red, and blue respectively.}
\label{fig:times240}
\end{figure}
\subsection{Computation times}
In Fig.~\ref{fig:times240} we plot the computation time evolution for \ac{ASP} and CellNOpt with respect 
to the number of experimental observations (i.e. output measures) included in the \textit{in silico} datasets used to run the optimizations. Since we generated multiple datasets which contained the same number of experimental observations, for each optimization related to these multiple datasets we obtained minimum, maximum, and average times. We observe that the \ac{ASP} computation times are in a range that goes from 0.02 to 0.15 seconds, while CellNOpt computation times to find the best score goes from 43 minutes to 2.7 hours, which was set as the limit running time. We see from these results that \ac{ASP} outperforms CellNOpt in 5 order of magnitude guaranteeing in all cases global optimality. As discussed in a previous subsection, the main prospective issue is to test the relevance of this conclusion when optimizing with real data instead of \textit{in silico} data. 


\section{Conclusion}
We have proposed a formal encoding of a combinatorial optimization problem related to the inference of Boolean rules describing protein signaling networks. We have used \ac{ASP}, a declarative problem solving paradigm, to solve this optimization problem and compared its performance against the stochastic method implemented by CellNOpt. Our ASP formulation relies on powerful, state-of-the-art and free open source software  [20,12]. As main conclusion, we prove that our \ac{ASP}-based approach ensures to find all optimal models by reasoning over the complete solution space. Moreover, in the experiments presented in this work, \ac{ASP} outperforms CellNOpt in up to 5 orders of magnitude.


Our analyses provide concrete illustrations of the potential applications, in our opinion under-explored, of \ac{ASP} in this field. Recently, \ac{ILP} have been used to solve the same problem that we described here~\cite{Mitsos2009}. In principle, \ac{ILP} solvers can also provide the complete set of optimal solutions but a detailed comparison between \ac{ASP} and \ac{ILP} for this particular problem remains to be done.

As discussed within the results section, several prospective issues shall now be investigated.  We first have to study the robustness of our results when optimizing over real networks and datasets. Second, we shall develop tools to explore the topology of the space of suboptimal models in order to gain in biological relevance in the inference process and try to elucidate whether this topology informs about the biological relevance of minimal models. Finally, by considering the presence of feedbacks loops in the input \ac{PKN} and by studying the effect of different discretization approaches, we hope to improve the state of the art in protein signaling network inference and offer a useful tool for biologists.

\section*{Acknowledgements}
\small
The work of the first author was supported by the project ANR-10-BLANC-0218. 
The work of the second author was financed by the BMBF MEDSYS 0315401B.
The work of the third author was partially supported by the 'Borsa Gini' scholarship, awarded by 'Fondazione Aldo Gini', Padova, Italy.

\bibliographystyle{splncs}

\newpage
\section*{Author contributions}
\small
JSR and CG started the project. SV proposed the logic formalization and built the ASP encoding, supported by ST. SV performed ASP experimentations, advised and coordinated by AS. FE conducted the benchmark generation and performed CNO experimentations, advised and coordinated by JSR and CG. Comparisons were coordinated by CG. All were involved in the writing. 

\end{document}